\documentclass[%
 notitlepage, 10pt, twocolumn,
 amsmath,amssymb,
 aps,prx,
floatfix,
]{revtex4-2}

\usepackage{physics}
\usepackage{amsthm}
\usepackage{bm}
\usepackage{graphicx}
\usepackage{dcolumn}
\usepackage{hyperref}
\usepackage[linesnumbered,boxed]{algorithm2e}
\usepackage{subfig}
\usepackage{mathtools}
\usepackage{xcolor}

\newcommand{\f}[2]{\dfrac{#1}{#2}} 
\newcommand{\p}[1]{\left(#1\right)}
\renewcommand{\sp}[1]{\left[#1\right]} 
\newcommand{\B}{\mathcal{B}}

\renewcommand{\sp}[1]{\left[#1\right]} 

\begin{document}

\preprint{APS/123-QED}
\title{Classical simulators as quantum error mitigators via circuit cutting}

\author{Ji Liu$^1$}
\email{ji.liu@anl.gov}

\author{Alvin Gonzales$^2$}
\email{agonza@siu.edu}

\author{Daniel Dilley$^1$}
\email{ddilley@anl.gov}

\author{Zain H. Saleem$^1$}
\email{zsaleem@anl.gov}

\affiliation{Mathematics and Computer Science Division, 
Argonne National Laboratory, 9700 S Cass Ave, Lemont IL 60439$^1$}
\affiliation{Intelligence Community Postdoctoral Research Fellowship Program, Argonne National Laboratory, Lemont, IL, USA$^2$}

\date{\today}

\begin{abstract}
We introduce a circuit cutting-based error mitigation framework, simulated quantum error mitigation (SQEM), in which an auxiliary error-mitigation subcircuit is simulated classically while the main computation is executed on quantum hardware. For the class of circuits considered here, SQEM can be implemented with polynomial overhead by mitigating one qubit at a time and classically recombining the resulting output distributions. This qubit-wise strategy avoids the exponential overhead of cutting all protected qubits simultaneously, while providing a scalable approximation to joint mitigation. We demonstrate the method using Pauli check sandwiching within VQE hardware-efficient ansätze and obtain ground-state energies close to noiseless ansatz simulations, including chemical accuracy for LiH.
\end{abstract}

\maketitle


\section{Introduction}
Currently available quantum hardware do not have the capability to perform fully fault tolerant quantum computation due to the non-availability of high fidelity qubits. However, due to the inherent decoherence in quantum mechanics from interactions of the system with the environment, errors are inevitable and any quantum computation is affected by noise. Various error mitigation techniques have been developed that do not require the large quantum overhead necessary in fault tolerant quantum error correction. Some examples are: zero noise extrapolation \cite{Temme2017_ErrorMitigForShorDepthQuantCirc, 1805.04492, tudor2020_digitalZeroNoiseExtrapForQuantErrMitig}, probabilistic error cancellation \cite{Temme2017_ErrorMitigForShorDepthQuantCirc}, readout error mitigation \cite{Nachman_2020ReadoutNoise}, Pauli check sandwiching (PCS) \cite{Debroy_2020ExtFlagGadgetsForLowOverCircVer, gonzales2022quantum}, and symmetry verification \cite{Bonet-Monroig2018_LowCostErrMitigBySymmVerif, McArdle2019_ErrMitigDigitalQuantSimulation, shaydulin2021error, Cai2021quantumErrorMitigSymmExpan}.

\begin{figure}[h]
\centering
\includegraphics[width= \linewidth]{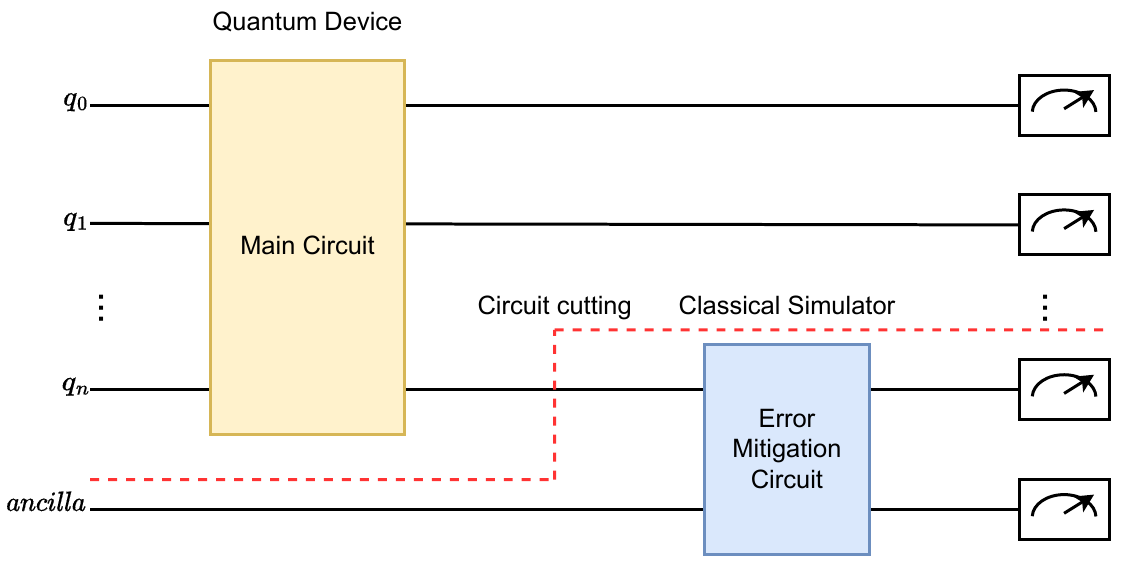}
\caption{Circuit cutting framework for quantum error mitigation}
\label{fig:framework}
\end{figure}

Many of these quantum error mitigation techniques insert additional gates and ancillas that need to be executed with the main circuit. One of the drawbacks of these techniques is that these error mitigation subcircuits introduce noise associated with the additional gates and ancillas that they require to run alongside the main circuit on the quantum hardware. In this work, we introduce the Simulated Quantum Error Mitigation (SQEM) framework, where the error mitigation subcircuits are run on a classical simulator whereas the main circuit runs on a quantum device. Essentially, our framework converts error mitigation techniques that require additional gates and ancillas almost entirely into classical pre and post processing procedures and thus makes the execution of the error mitigation protocol almost noiseless.

The SQEM framework relies on the quantum circuit cutting also known as ``quantum knitting". The quantum circuit cutting techniques allow us to divide large quantum circuits into smaller fragments~\cite{peng2020simulating, perlin2021quantum,  ayral2021quantum,ayral2020quantum, tang2021cutqc}. These fragments can then be run on separate quantum devices and then the output distributions from each of these fragments can be combined to generate the output of the full quantum circuit. In our error mitigation framework, we fragment the quantum circuit into two parts: the main/original quantum circuit which is executed on a quantum device and the quantum error mitigation subcircuits which are sent to a classical simulator. This is demonstrated in Fig.~\ref{fig:framework}. This allows us to error mitigate almost noise free. Since the error mitigation sub-circuit is run noiselessly on a simulator, it allows us to detect all kinds of errors including the non-Markovian errors. The state preparation and measurement (SPAM) errors due to circuit cutting is also mitigated by the mitigating circuit. In our demonstration on quantum hardware, the error mitigation method that we use is Pauli check sandwiching but other error mitigation schemes can also be used. 

One of the objections often raised with the circuit cutting technique is that the computational cost is exponential in the number of cuts. Therefore, if we were to error mitigate all the qubits at once we will have to incur an exponential computational cost. Our framework avoids this scalability problem because we error mitigate one qubit at a time. We provide examples of quantum circuits where this is possible. Each of the instances where we error mitigate a qubit generates a probability distribution. 
We introduce a method for recombining these different probability distributions such that the re-combined probability distribution is error mitigated from the noise across all qubits. We demonstrate the efficacy of our framework on quantum hardware. Specifically, we perform three variational quantum eigensolver (VQE) demonstrations to estimate the ground state energies for the molecules LiH, BeH$_2$, and HF. Our results match the noiseless ansatz result to a very high accuracy as shown in Figs.~\ref{subfig:results_lih} and \ref{subfig:results_HF}.

SQEM is most natural when the mitigation routine can be written as a small auxiliary subcircuit attached to the main computation through only a few wires. This includes cases such as PCS, symmetry channel verification \cite{tsubouchi_2025}, and virtual distillation~\cite{li2023enhancing}, where the checking or postselection operations can be separated from the main circuit and treated as their own fragment. In this work, we focus on a qubit-wise version of this idea, which keeps the number of cuts per run small and avoids the exponential overhead of cutting all protected qubits at once. This qubit-wise strategy is a scalable approximation, not a general exact correction method for arbitrary strongly correlated multi-qubit noise.

\section{SQEM framework}

Simulated quantum error mitigation (SQEM) transforms the error mitigation components of an input circuit into classical pre- and post-processing procedures by applying circuit cutting. The fragments that perform error mitigation are sent to a classical computer to be simulated and the main circuit is sent to the quantum hardware. This procedure enables the gates that comprise the error mitigation scheme to be noiselessly executed, while the original unmitigated quantum computation is run on quantum hardware. Fig~\ref{fig:framework} provides a schematic view of this protocol. 

The validity of SQEM follows from the circuit-cutting reconstruction: once the full circuit is decomposed into a main fragment and an error-mitigation fragment, the output of the original circuit can be recovered from the outputs of the fragments through classical post-processing. In SQEM, the mitigation fragment is simulated classically rather than executed on quantum hardware, so the target reconstructed distribution is unchanged in the ideal cutting procedure, while the additional hardware noise from the mitigation subcircuit is avoided. The computational advantage of our approach comes from applying this procedure qubit by qubit, which keeps the number of cuts per run small and avoids the exponential overhead associated with protecting all qubits simultaneously.

\subsection{Quantum Circuit Cutting}

A quantum circuit can be represented as a graph. In Ref.~\cite{peng2020simulating} it was shown how one can fragment such a graph into two or more subgraphs or subcircuits. These subgraphs can be cut by performing a complete set of measurements on the output of one subcircuit at a cut, and then preparing a corresponding set of states on the input of the other subcircuit at that cut. In practice, one can independently execute and measure the outputs of these subcircuits and classically post-process these results to obtain the output of the original uncut quantum circuit \cite{perlin2021quantum}.

Circuit cutting works by resolving an $n$-qubit quantum state $\rho$ and inserting a set of complete measurement basis, 
\begin{equation}
  \rho \simeq \f12 \sum_{M\in\B} M \otimes \tr_n\p{M_{(n)}\rho},
  \label{eq:cut_identity}
\end{equation}
where $\simeq$ means equality up to a permutation of the qubit order; $\B$ is a basis of Hermitian $2\times 2$ matrices with normalization $\tr\sp{M^{(i)} M^{(j)}} = 2 \delta_{ij}$ for $M^{(i)},M^{(j)}\in\B$; $\tr_n$ is the partial trace of qubit $n$; and $M_{(n)}$ is an operator that acts with $M$ on qubit $n$ and with identity on all the other qubits.
We use the set of Pauli operators plus the identity operator, $\B\equiv\{I,X,Y,Z\}$, as our basis. 

SQEM is most naturally applicable when the mitigation routine can be expressed as a small auxiliary subcircuit appended to, or interfaced with, the main computation through only a few wires. Examples include PCS and symmetry-verification settings in which ancillary checking or postselection operations form a separable verification block. In this work, we focus on a qubit-wise realization of SQEM, which keeps the number of cuts per run small and thereby avoids the exponential overhead of cutting all protected qubits simultaneously. This qubit-wise strategy is a scalable approximation and is not intended as a general exact correction method for arbitrary strongly correlated multi-qubit noise.

\subsection{Pauli Check Sandwiching}
Pauli Check Sandwiching~\cite{gonzales2022quantum} (PCS) uses pairs of checks to error mitigate the input circuit $U$ \cite{Debroy_2020ExtFlagGadgetsForLowOverCircVer,gonzales2022quantum}. Each pair of check sandwiches $U$ and leaves it invariant. 
We use a set of $n$ pairs of unitaries $\{\tilde{C}_1^{(i)}, \tilde{C}^{(i)}_2\}$, where $i$ represents the $i^\text{th}$ pair and each pair satisfies the condition
\begin{align}\label{eq:conditionOnChecks1}
    \tilde{C}_2^{(i)}U\tilde{C}^{(i)}_1=U.
\end{align}
Then the checks for the circuit are given by the set of pairs of controlled unitaries $\{C_1^{(i)}, C_2^{(i)}\}$ 
\begin{align}
    C_1^{(i)}&=\tilde{C}^{(i)}_1\otimes\op{-}_{i}+\mathbb{I}\otimes\op{+}_{i}\\
    C_2^{(i)}&=\tilde{C}^{(i)}_2\otimes\op{-}_{i}+\mathbb{I}\otimes\op{+}_{i},
\end{align}
where the control is on the $i^{th}$ ancilla and the target are the compute qubits. The error mitigation scheme uses the circuit $C_2^{(n)}\cdots C_2^{(2)}C_2^{(1)}UC_1^{(1)}C_1^{(2)}\cdots C_1^{(n)}$ and post selects on the zero outcome for a $Z$ basis measurement on the ancillas. In the ideal case that the error mitigation part of the circuit is noiseless and given arbitrary noise on the compute qubits, there always exists a finite number of checks such that the post selected state is noiseless.x

\begin{figure*}[htbp]
\centering
\includegraphics[width=\textwidth]{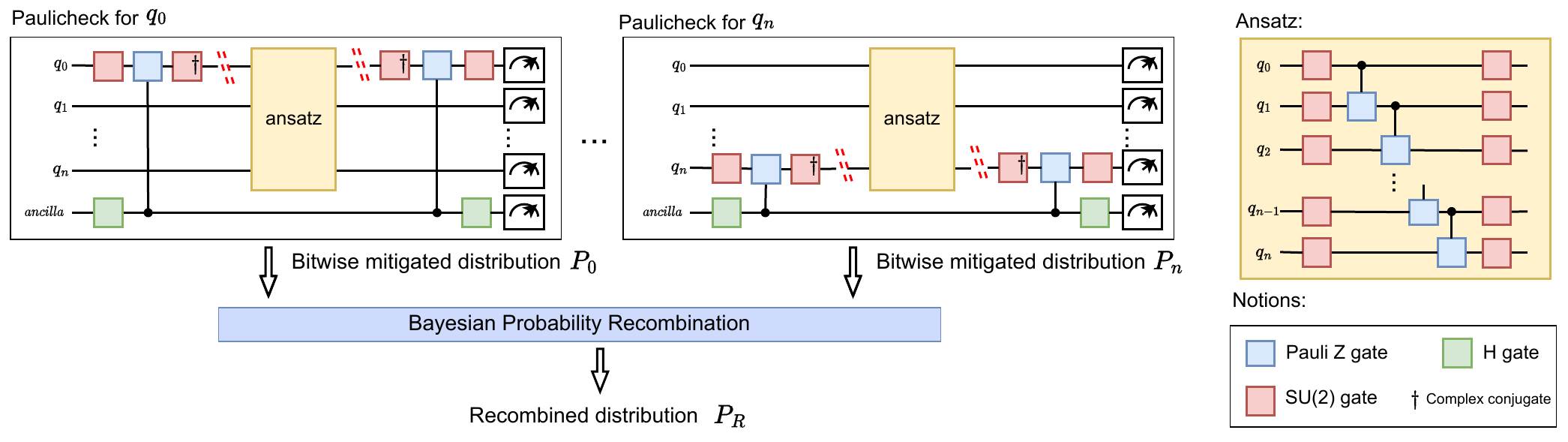}
\caption{Overview of the SQEM framework. The red blocks and the red blocks with a dagger in the PCS circuit are complex conjugates of each other. The framework involves three steps: 1) apply the error mitigation circuit for each individual qubit or a subset of qubits 2) cut out the error mitigation circuit fragment and run the original fragment and the mitigation fragment separately on a quantum device and a classical device. 3) recombine a probability distribution $P_R$ based on individual error mitigated distributions $P_0$ ... $P_n$ and a Bayesian probability recombination algorithm.  }
\label{fig:SQEM_overview}
\end{figure*}

For this research, we apply the PCS scheme on a hardware-efficient VQE ansatz, which consists of two layers of SU(2) gates and a cascade of controlled-Z gates. The check pairs we use are Z-gates with extra single-qubit gates and we check only a single qubit at a time as shown in Fig.~\ref{fig:SQEM_overview}. The single qubit check Pauli-Z checks are possible because they commute with the ansatz along with the SU(2) gates. The Pauli-Z checks remove all the Pauli-X and Pauli-Y errors. 


In PCS, the post selection probability exponentially decreases as the number of pairs of checks increases~\cite{gonzales2022quantum}. Our strategy resolves this problem by error mitigating one qubit at a time. With our strategy the total overhead in the number of runs required is only a polynomial function of the number of pairs of checks used. No single run directly produces a fully mitigated global output distribution because each run protects only one qubit. We therefore combine the information from the qubit-wise protected runs in a separate recombination step.

\subsection{Probabilities Recombination}

Applying SQEM one qubit at a time keeps each run small and avoids the exponential cost of protecting all qubits at once. The tradeoff is that no single run directly gives a fully mitigated output distribution, so the results from the qubit-wise protected runs must be combined afterward. Here we use a Bayesian recombination model to recombine the probability distributions. The Bayesian recombination model is analogous to Bayesian updating~\cite{berger1994bayesianupdate} in statistics. It is developed based on the assumption that the protected/checked qubits' probability distribution has less noise than unprotected/unchecked ones. This qubit-wise strategy does not retain the full ideal guarantee of joint PCS, but replaces it with a scalable approximation: each protected run is assumed to yield a less noisy bitwise marginal, and these marginals are recombined to improve the full output distribution. In this work, the effectiveness of this approximation is supported empirically by the hardware results on ibm$\_$auckland.

Considering an n-qubit circuit, we use PCS with SQEM to check each qubit individually. For the n runs we will get n distributions $P_{M\_1}$, ... , $P_{M\_n}$. $P_{UM}$ and $P_{M\_k}$ are the output probability distributions of the unmitigated runs and that of the mitigated run which protects qubit $k$ respectively. Each classical state $\ket{x_1x_2...x_n}$ in the computational basis is represented as a binary string $s = [x_1, x_2, ..., x_n]$, $s \in \mathbb{B}_c$, where $\mathbb{B}_c$ is the computational basis. $s[k] = x_k$, $k \in [0, n]$ indicates the measurement result of the $k$th qubit in the string $s$.  $P_{UM}(s)$ is the probability of getting the output $s$. For a distribution $P$, the probability of qubit $k$ having a measured result $\ket{0}$ over all the states is denoted as
\begin{equation}
  w_0(P, k) =  \sum_{s\in \mathbb{B}_c} P(s)\text{, if } s[k] = 0.
  \label{eq:uw0}
\end{equation}
Similarly, we have the probability $w_1(P, k)$ for outcome $\ket{1}$. $w_0(P_{M\_k}, k)$ is considered as the ideal bitwise probability for qubit $k$ with outcome $\ket{0}$ since we use the probability from the run that protects qubit $k$. We initialize the recombined probability $P_R$ to $P_{UM}$ and update $P_R$ with the update probabiltities $P_{update}$
\begin{equation}
P_{R}'(s) = P_{R}(s) + P_{update}(s), \forall s\in \mathbb{B}_c.
  \label{eq:p_r}
\end{equation}

The probability $P_{update}(s)$ updates the recombined probability distribution $P_R$ such that it has bitwise probabilities close to the protected bitwise probabilities in each mitigated result. For each string $s$, the update probability is calculated based on the gradient of the protected distribution $P_{M\_k}$ for each qubit $k$ and the current distribution $P_R$
\begin{equation}
  \resizebox{0.88\hsize}{!}{
 $ P_{update}(s) = \sum_{k\in [0,...,n]} P_{R}(s)\times \frac{w_j(P_{M\_k}, k)}{w_j(P_R, k)}, j = s[k].$
  }
  \label{eq:p_update}
\end{equation}

\begin{figure*}[htbp]
\centering
\subfloat[LiH interatomic distance $l = 1.5$\AA\label{subfig:results_lih}]{\includegraphics[width=0.49\linewidth]{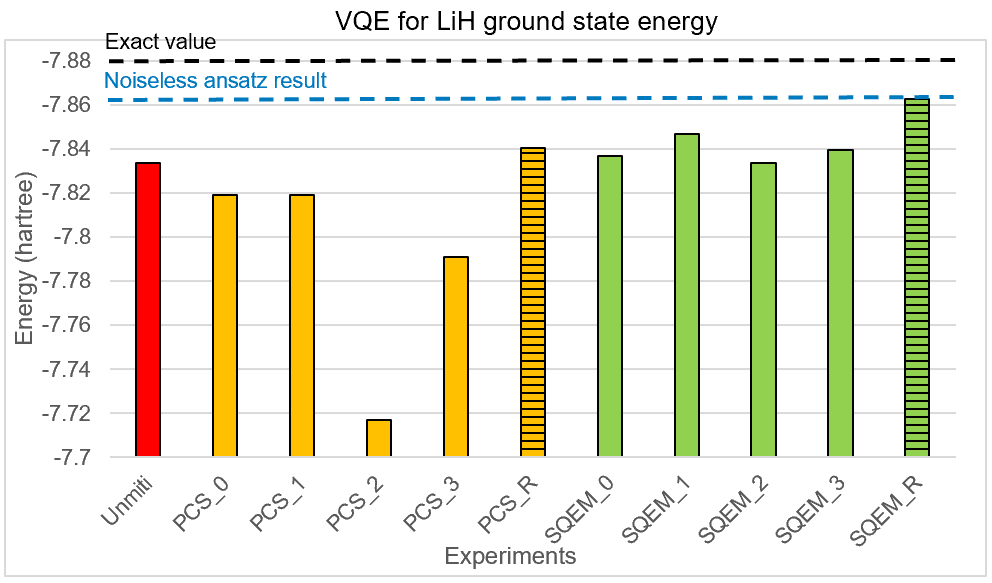}}
\subfloat[HF interatomic distance $l = 0.9$\AA\label{subfig:results_HF}]{\includegraphics[width=0.49\linewidth]{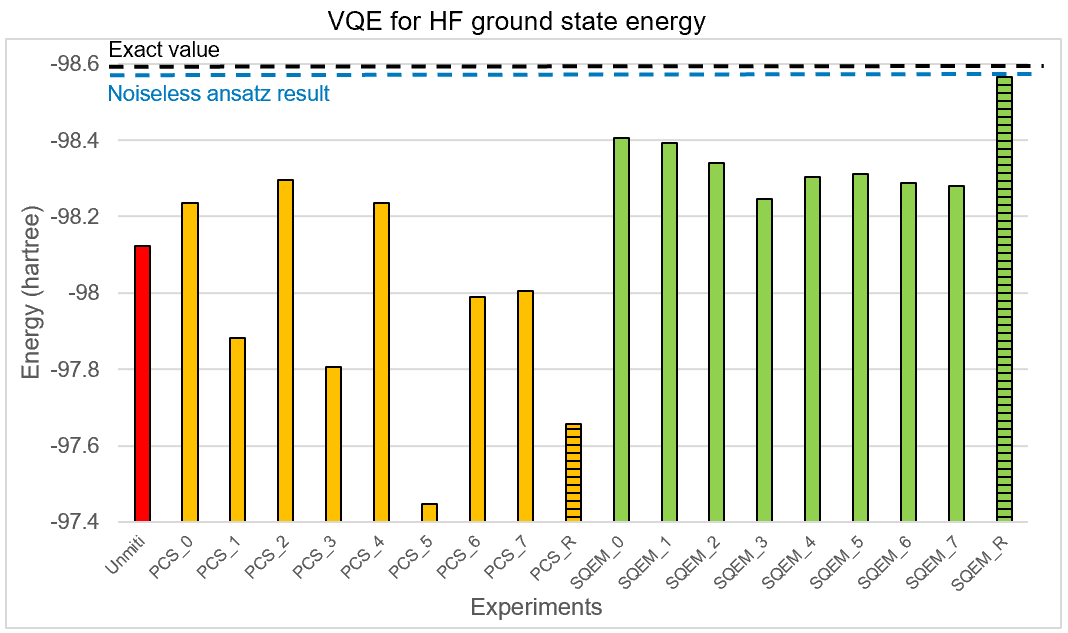}}
\caption{Quantum hardware demonstration solving LiH and HF ground state energy with VQE. The red bar represents the unmitigated result. The orange bars (PCS\_k) are the results from the runs with bitwise PCS on real device. The green bars (SQEM\_k) represent the results that apply SQEM to qubit $k$. The orange and green bar with horizontal stripes represent the result by Bayesian recombining the distributions from the orange and the green bars.}
\label{fig:results}
\end{figure*}

The recombined probability $P_R$ is iteratively updated and renormalized. The algorithm stops when the hellinger distance~\cite{hellinger1909hellinger} between $P_{R'}$ and $P_{R}$ is smaller than a threshold. After the process, the recombined probability distribution $P_{R}$ has bitwise probabilities close to the mitigated bitwise probabilities:
\begin{equation}
 w_j(P_R, k) \simeq  w_j(P_{M\_k}, k), \forall j\in[0,1], k\in[0,n].
  \label{eq:p_eq}
\end{equation}
In practice, the recombination is carried out on the observed output bitstrings obtained from the experiment rather than on the full probability vector over all computational basis states. Therefore, the amount of data updated during recombination is upper bounded by the number of sampled outcomes, which is at most the total number of shots.



\section{Hardware Demonstration}
We apply SQEM to the Variational Quantum Eigensolver (VQE) task of ground state energy estimation for three molecules: LiH, BeH$_2$, and HF. The number of qubits in the circuits are 4, 6, and 8 for LiH, BeH$_2$, and HF, respectively. We test our error mitigation framework with the two-local SU(2) ansatz~\cite{hardware-efficient} on the 27-qubit quantum device \texttt{ibm\_auckland}. 
The Hamiltonians are constructed in the STO-3G basis with frozen orbitals. The ansatz is initialized with Hartree-Fock~\cite{hartreefork} initialization. We fixed the parameters to the optimal parameters found with the `COBYLA' optimizer on a noiseless simulator. The Bayesian recombination threshold is set to 0.0001. Each run is executed with 10000 shots. The code used for demonstrations in this work is available online at \url{https://github.com/revilooliver/cut4mitigation}.

The blue dashed lines in Figure~\ref{fig:results} represent the noiseless results of the ansatz. They are derived from a noiseless simulator with the same number of shots as our hardware runs. The black dashed lines represent the exact ground state energy calculated from PySCF~\cite{sun2018pyscf} and Qiskit Nature~\cite{Qiskit}. Figure~\ref{subfig:results_lih} shows the results for LiH. The unmitigated result is -7.8334 hartree which is far from the noiseless ansatz result -7.8634 and the exact gound state energy -7.8810. Then, we run the ansatz with bitwise PCS on a real device and the results are marked with orange bars in the figure. Since the error mitigation circuit introduces extra noise, the results may get worse. 

We apply SQEM to simulate the error mitigation fragment on a classical device. As shown in the figure, the error mitigated results (greed bars) with SQEM are closer to the ideal results than the orange bars. At last we use the Bayesian recombination algorithm to recombine the probability distributions from the green bars. The recombined result has energy of -7.8625. It is within the chemical accuracy (an energy error of approximately 0.0016 hartree) of the noiseless ansatz result -7.8634. The results show that SQEM can significantly mitigate the noise on quantum devices and generate results close enough to the results on a noise-free simulator.

We also include the results for HF in Figure~\ref{subfig:results_HF}. The exact groud state energy is -98.5929 hartree. The unmitigated result energy is -98.1218. The recombined result energy is -98.5653, which is very close to the noiseless ansatz result -98.5684. For the BeH$_2$ molecule, the unmitigated, recombined, and noiseless results are -15.3770, -15.4678, and -15.4706, respectively. 

\section{Conclusions and Future Directions}

We introduce the SQEM framework which removes the added gates and ancillas required by many error mitigation techniques from the quantum hardware execution by applying circuit cutting. The error mitigation subcircuit is executed on a classical device whereas the main circuit is executed on a quantum device. Furthermore, we demonstrate that this can be performed in a polynomially scalable way for a variety of problems by error mitigating individual qubits with PCS and performing a Bayesian recombination of the probability distributions at the end. 

While we used Pauli check sandwiching, our framework is compatible with other error mitigation techniques whose mitigation routine can be isolated as an auxiliary subcircuit and reconstructed through circuit cutting. As a concrete example, SQEM can be combined with symmetry verification. In symmetry-verification protocols, ancilla-assisted checks of conserved symmetries are appended to the main circuit and followed by measurement and postselection. When the verification module can be isolated as a small auxiliary subcircuit connected to the computational register through only a few wires, SQEM can be applied by cutting those connections. The main computational fragment is then executed on hardware, while the symmetry-check fragment is simulated classically, and the symmetry-verified output is reconstructed through standard circuit-cutting post-processing.

In this way, SQEM preserves the ideal action of symmetry verification while avoiding the additional hardware noise introduced by the ancilla, controlled operations, and measurements used in the verification routine. By contrast, methods such as randomized compiling do not naturally fit this structure since their mitigation mechanism is not given by a separable auxiliary circuit fragment. We tested our ideas with three hardware VQE demonstrations: estimating the ground state energies of LiH, BeH$_2$, and HF. In all three demonstrations, the SQEM framework matched the results of a full classical simulation of the VQE circuits to a very high accuracy.

For future directions, we can look at applying SQEM for other problems such as the quantum approximate optimization algorithm \cite{farhi2014QAOA}, investigating the effects of protecting multiple qubits at a time, improving the probability recombination algorithm, testing larger molecules, and trying other error mitigation schemes.

\section*{Acknowledgments}
We thank Mark Byrd, Paul Hovland, and Gokul Subramanian Ravi for helpful discussions. 
This material is based upon work supported by the U.S. Department of Energy Office of Science National Quantum Information Science Research Centers. This research was also supported in part by an appointment to the Intelligence Community Postdoctoral Research Fellowship Program at Argonne National Laboratory, administered by Oak Ridge Institute for Science and Education through an interagency agreement between the U.S. This research used resources of the Oak Ridge Leadership Computing Facility, which is a DOE Office of Science User Facility supported under Contract DE-AC05-00OR22725."

The submitted manuscript has been created by UChicago Argonne, LLC, Operator of Argonne National Laboratory (``Argonne”). Argonne, a U.S. Department of Energy Office of Science laboratory, is operated under Contract No. DE-AC02-06CH11357. The U.S. Government retains for itself, and others acting on its behalf, a paid-up nonexclusive, irrevocable worldwide license in said article to reproduce, prepare derivative works, distribute copies to the public, and perform publicly and display publicly, by or on behalf of the Government. The Department of Energy will provide public access to these results of federally sponsored research in accordance with the DOE Public Access Plan (\url{http://energy.gov/downloads/doe-public-access-plan}).

\bibliography{references}

\end{document}